# Electronic transport through a silicene-based zigzag and armchair junction

Dace Zha[a]   Changpeng Chen[a,b]   Jinping Wu[a,b]

a. School of science, Wuhan University of Technology, Wuhan 430070 PR China

b. Material Science and Chemistry Engineering College, China University of Geosciences, Wuhan, 430074, P. R. China

**Abstract:** Using density functional theory and non-equilibrium Green's function technique, we performed theoretical investigations on the transport properties of several ZAZ SiNRs junctions (a similar kind of silicene molecules junction combined by zigzag and armchair silicene nanoribbons). It is found that the differential conductances of the three systems decrease with an order of 5-ZAZ>4-ZAZ>3-ZAZ. Particularly, the Negative differential resistance (NDR) can be observed within certain bias voltage range only in 3-ZAZ SiNRs. In order to elucidate the mechanism the NDR behavior, the transmission spectra and molecular projected self-consistent Hamiltonian (MPSH) states are discussed in details.

## 1. Introduction

During the past decades, progress in micro-fabrication and self-assembly techniques [1] has made it possible to design molecular devices. A number of molecular devices have shown their peculiar current–voltage (I–V) properties, such as negative differential resistance(NDR) [2,3], rectification [4], and current switch [3,5], etc. On the other hand, recent advances in controllable synthesis and characterization of nanoscale materials, have opened up important possibilities for the investigation of ultra-thin 2D systems. The nanoribbons of graphene, MoS2, BN etc. are found to possess variety of unusual properties which can be utilized in molecular devices [6,7,8]. The excellent research success of graphene has led to the investigation of other honeycomb monolayers composed of group IV elements, such as silicene, germanene, etc. Silicon, as an element resides in the same column of the periodic table with carbon, has recently attracted increasingly concern for its existence of a honeycomb monolayer [9]. As is well known, Silicon owns four valence electrons as carbon does, having many similar characters to carbon. However, Si also possesses some unique properties that carbon hasn't. Recent theoretical and experimental studies have shown that different from graphene, silicene is not planar but has a low-buckled honeycomb structure with buckled height of about 0.44A for stability [10,11,12]. The underlying reason is that silicene's sp3 hybridization is more stable than its sp2 form [13,14]. Similar to graphene nanoribbons (GNRs), silicene nanoribbons(SiNRs) also have two kinds of shapes: zigzag and armchair. Studies show that armchair SiNRs exhibit semiconductor behavior while zigzag SiNRs have stable anti ferromagnetic states [15,16,17,18,19]. The hydrogen-terminated GNR with either armchair- or zigzag-shaped edges has an energy gap which can be tuned by width or edge shape and shows interesting 3n rule, which was shown theoretically by Son et al. [20] and confirmed experimentally by Li et al. [21] and Ritter et al. [22]. As is well known, a transistor can potentially be made by using zigzag nanoribbon as source and drain and a semiconductor armchair nanoribbon as the channel on the same graphene layer [23,24]. So, it is naturally for one to consider a similar kind of silicene molecules junction combined by zigzag and armchair silicene nanoribbons, which is called ZAZ SiNRs junction (shown in Fig.1).

In this present work, we study the electron transmission through the connections between zigzag SiNRs and armchair SiNRs using the nonequilibrium Green's function (NEGF) methodology. It is found that some interesting properties exist in certain kind of N-ZAZ SiNRs, such as negative differential resistance (NDR), which offer important insights into the potential application of SiNRs as both transistors and interconnects.

## 2. Model and method

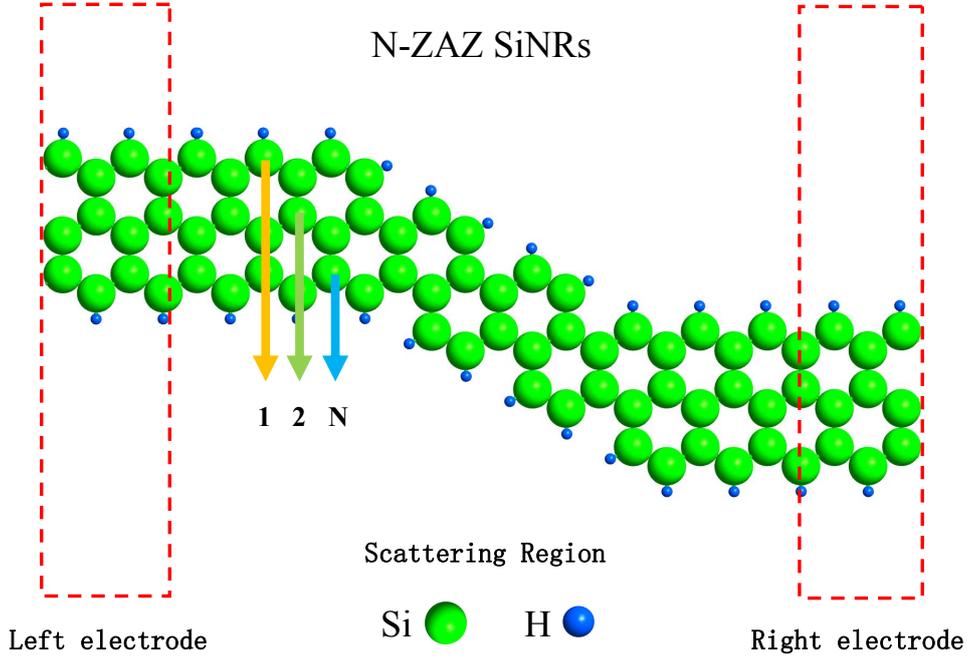

Fig. 1. The structure of the two-probe system of N-ZAZ SiNRs from top view. The N-ZAZ SiNRs are named after the number of the dimer lines across the ribbons width for the ZAZ SiNRs edges.

The optimized geometric structure of ZAZ SiNRs ~~from the top view~~ are illustrated in Fig. 1. The N-ZAZ SiNRs are named after the number of the dimer lines across the ribbons width for the ZAZ SiNRs edges. The structures were optimized and the quantum transport calculations were carried out by the Atomistix ToolKit (ATK) package [25-26], which bases on the fully self-consistent non-equilibrium Green's functions and the density functional theory. Under external bias, the current through a molecular junction is calculated from the Landauer–Büttiker formula:

$$I(V) = 2e/h \int [f_L(E-\mu_L) - f_R(E-\mu_R)] T(E,V) dE$$

Where $f_{L/R}$ is the Fermi–Dirac distribution for the left (L) and right (R) electrodes, $\mu_{L/R}$ the electrochemical potential of the left (L) or right (R) electrodes and $T(E,V)$ is the transmission coefficient at energy $E$ and bias voltage $V$.

$$T(E,V) = Tr[\Gamma_L(E,V) G^R(E,V) \Gamma_R(E,V) G^A(E,V)]$$

where $G^{R/A}$ are the retarded and advanced Green's functions, and coupling functions $\Gamma_{L/R}$ are the imaginary parts of the left and right self-energies, respectively. Self energy depends on the surface Green's functions of the electrode regions and comes from the nearest-neighbor interaction between the extended molecule region and the electrodes. For the system at equilibrium, the conductance $G$ is evaluated by the transmission function $T(E)$ at the Fermi level (FL) $E_F$ of the system: $G=G_0 T(E_F)$, where $G_0=2e^2/h$ is the quantum unit of conductance, and $h$ is Planck's constant and $e$ is the electron charge.

The geometric optimization and transport properties are calculated by ATK within local density approximation for exchange-correlation potential. The exchange correlation functional is set as local-density approximation (LDA) in the PerdewZunger (PZ) form. To save computational time of the transport calculations, a mesh cutoff is set to be 150 Ry and the Monkhorst–Pack (1×1×100) K-point grid is used. The basis set is adopted for elements of systems and the convergence criterion is 10-4 Hartree for the total energy. Test calculations with the larger basis set size, larger cutoff energy,and dense K-point (i.e. 3×3×100) are also performed, which give similar Results.

## 3. The transport properties of N-ZAZ SiNRs

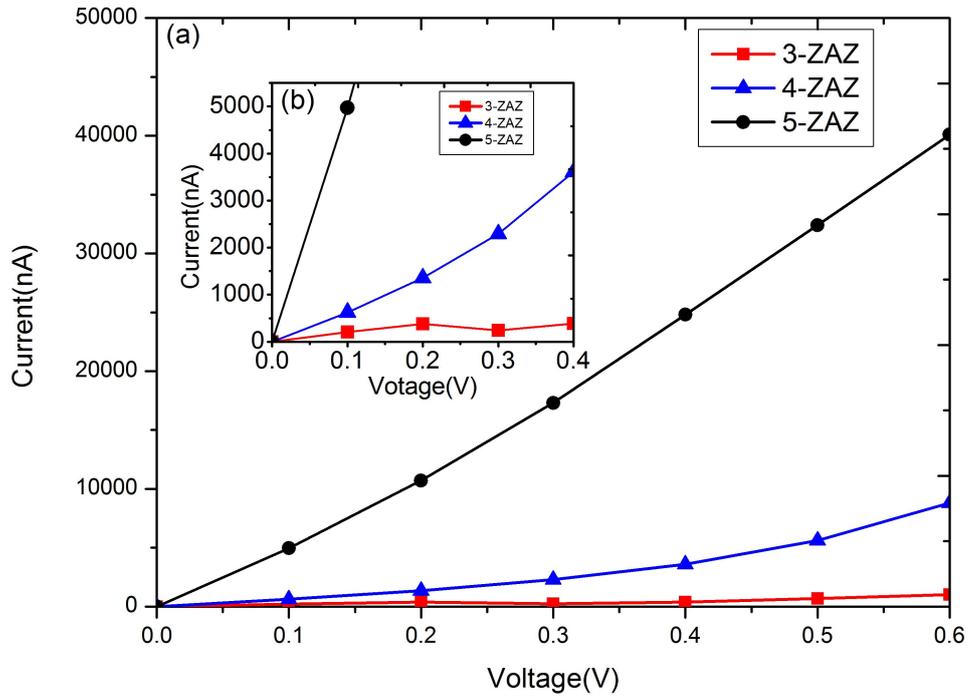

Fig. 2. The currents as functions of the applied bias for N-ZAZ SiNRs. The inset shows the local expansion for lower bias range.

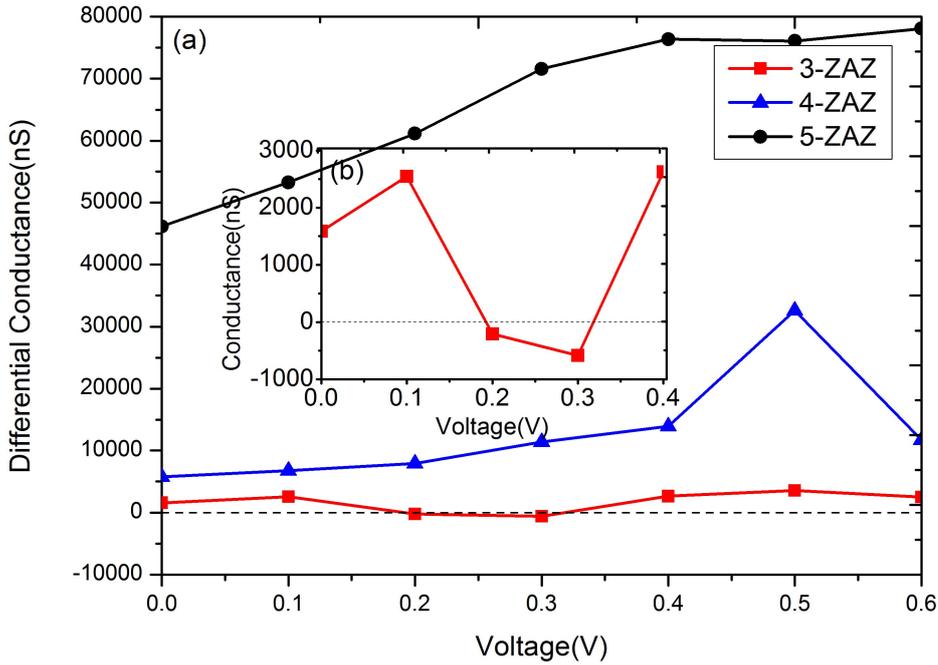

Fig. 3. Differential conductance–Voltage curves of N-ZAZ SiNRs. The insert shows the local expansion for lower bias range in 3-ZAZ SiNRs.

The I–V characteristics of 3-ZAZ SiNRs, 4-ZAZ SiNRs and 5-ZAZ SiNRs in the range of [0V, 0.6V] are calculated in Fig. 2. It can be found that the 5-ZAZ SiNRs show the metal behavior and the I-V curve displays a linear behavior at a bias voltage. The current through the three systems decrease with an order of 5-ZAZ > 4-ZAZ > 3-ZAZ. In particular, the current through 5-ZAZ is much higher than that through 3-ZAZ under the same applied bias. This means that 3-ZAZ can weaken the electron transport at the same bias voltage. In addition, when the bias takes a value between 0.1 V and 0.3 V, the current of 3-ZAZ SiNRs shows NDR behavior in the bias range. When V > 0.4 V, the currents rapidly increase with bias voltage. However, the currents of 4-ZAZ SiNRs and 5-ZAZ

SiNRs always increase with bias voltage. The NDR behaviors for 4-ZAZ SiNRs and 5-ZAZ SiNRs are not found. In Fig. 3(a), we plot the corresponding conductance (dI/dV) curves for 3-ZAZ SiNRs, 4-ZAZ SiNRs and 5-ZAZ SiNRs. Under zero bias, the conductance of 3-ZAZ SiNRs, 4-ZAZ SiNRs and 5-ZAZ SiNRs is $1.59\times10^3$nS, $5.57\times10^3$nS, and $4.62\times10^4$nS, respectively. The differential conductance of 4-ZAZ SiNRs and 5-ZAZ SiNRs first increases with the applied bias increasing, which means that the corresponding transmission channels are opened and make contributions to the electronic transport. However, as for 3-ZAZ structure, the differential conductance is depressed in comparison with that of 4-ZAZ and 5-ZAZ structure, which indicates that 3-ZAZ depress the transport properties. Especially, the negative value of conductance can be found in bias range of [0.2V, 0.3V] for 3-ZAZ SiNRs, which matches well with I-V curves for the explored bias region.

The characteristic of I–V depends on the transmission spectrum of zero bias when the external bias is very low. The transmission spectra T(E,V) of N-ZAZ SiNRs is calculated under zero bias voltage in Fig. 4. We can clearly see that there are strong transmissions for 4-ZAZ and 5-ZAZ structure. So the metal characteristic for 4-ZAZ and 5-ZAZ structure can be observed in Fig. 2(a). It is well known that the transmission coefficient is related to the wave function overlap between molecule and electrode, namely, the coupling degree between molecular orbitals and the incident states from the electrode. It can be found from Fig. 4 that four frontier molecular orbitals, namely, HOMO, HOMO-1, LUMO and LUMO+1 lie near the Fermi energy level, which indicates that the electron transport at low bias voltage is mainly contributed by the HOMO, HOMO-1, LUMO and LUMO+1. The average transmission coefficient of 5-ZAZ structure is higher than those of 3-ZAZ structure and 4-ZAZ structure, which indicates that the electron transport ability of 5-ZAZ SiNRs is obviously higher than 3-ZAZ SiNRs and 4-ZAZ SiNRs, so the current of 5-ZAZ SiNRs is remarkably higher than those of 3-ZAZ SiNRs and 4-ZAZ SiNRs.

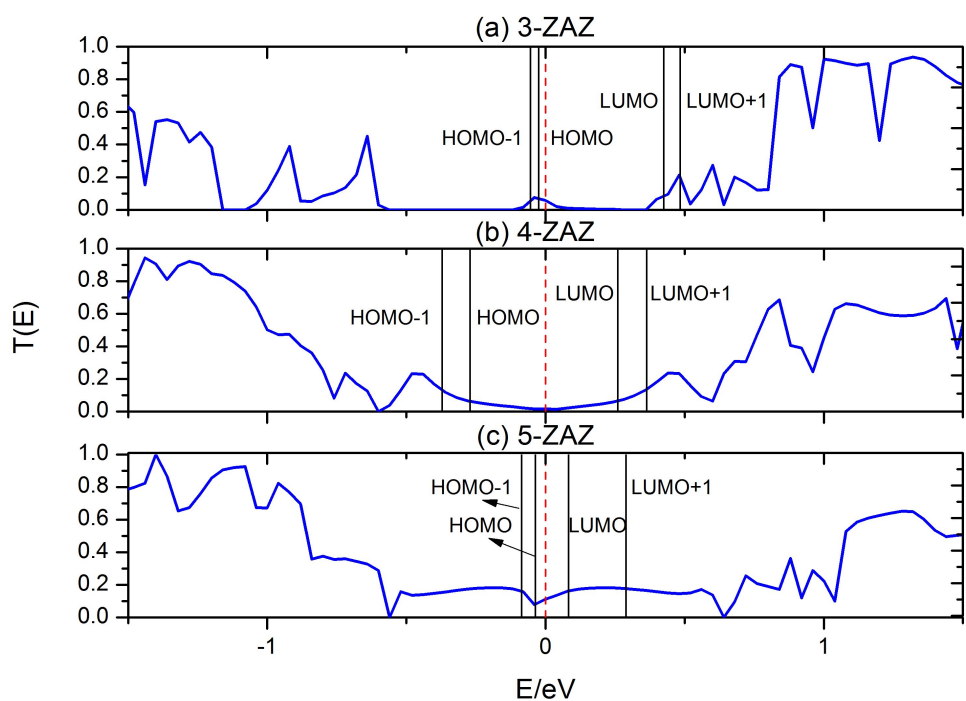

Fig. 4. The transmission spectra of N-ZAZ SiNRs under zero bias voltage. The vertical red dash line is the Fermi level.

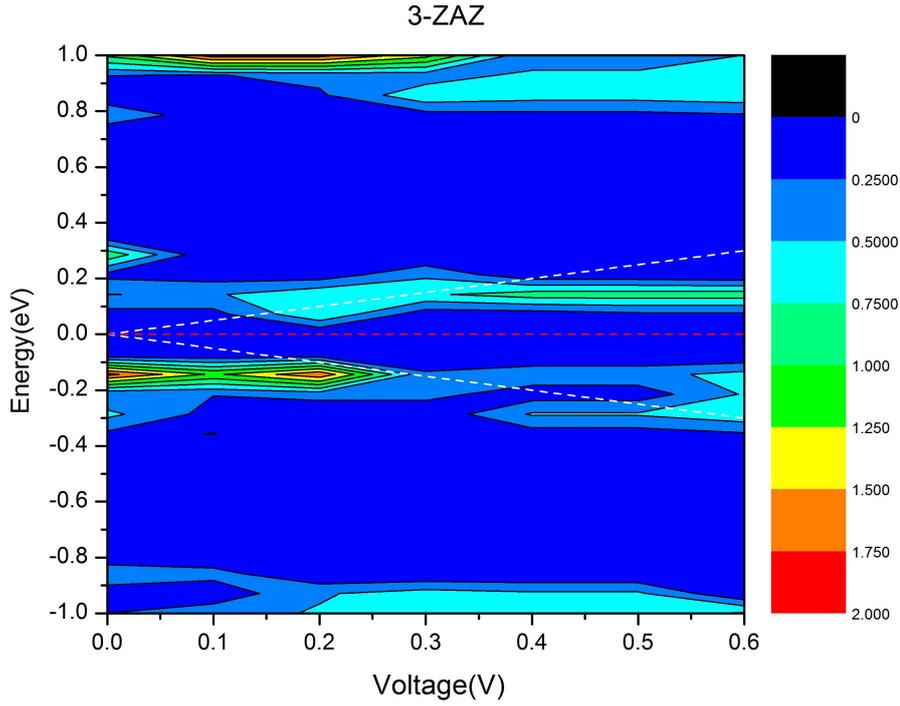

Fig. 5. The total Transmission Spectrum as a function of the bias voltage and electron energy for 3-ZAZ SiNRs.

To understand the mechanism responsible for NDR, we calculated the total transmission coefficient as a function of the bias and electron energy for 3-ZAZ SiNRs in Fig. 5. In this bias voltage-dependent transmission spectra, the bottom panel is bias voltage from 0 to 0.6 V, the left panel is Energy which is the electron energy from -1 to 1 eV around Fermi level and color shows the value of transmission coefficient, where red indicates the largest value and black means almost zero. The horizontal red dash line shows the average Fermi level that is the average value of chemical potential of the right and left electrodes. The two gradient white dash lines show bias window. The region between two dash lines is the bias window. From Fig. 5, we find that there are smaller transmission coefficients in bias windows at the lower bias range. With an increase in the bias, the transmission comes into bias windows which result in the increasing of currents. However, we can find the total magnitude of transmission coefficient coming into the bias window becomes smaller in the range from 0.2 V to 0.3 V. So the NDR behavior appears in 3-ZAZ SiNRs.

Choosing several typical bias voltage points, we calculate the bias voltage-dependent MPSH eigenvalue and eigenstate of scattering region combined with the transmission spectra to further understand the origin of the NDR phenomenon appearing in 3-ZAZ structure. Fig. 6 show the MPSH of the highest occupied molecular orbital (HOMO), the lowest unoccupied molecular orbital (LUMO), and their nearby orbitals HOMO−1 and LUMO+1. It can be seen that both HOMO-1, LUMO and LUMO+1 states are obviously delocalized for 3-ZAZ structure at 0.2 V. But LUMO is nearly localized for 0.3 V. Note that if the orbit is delocalized across the molecule, an electron entering the molecule at the energy of the orbital has a high probability of going through the molecule [27]. So we can observe the NDR phenomenon in I-V curve of 3-ZAZ structure.

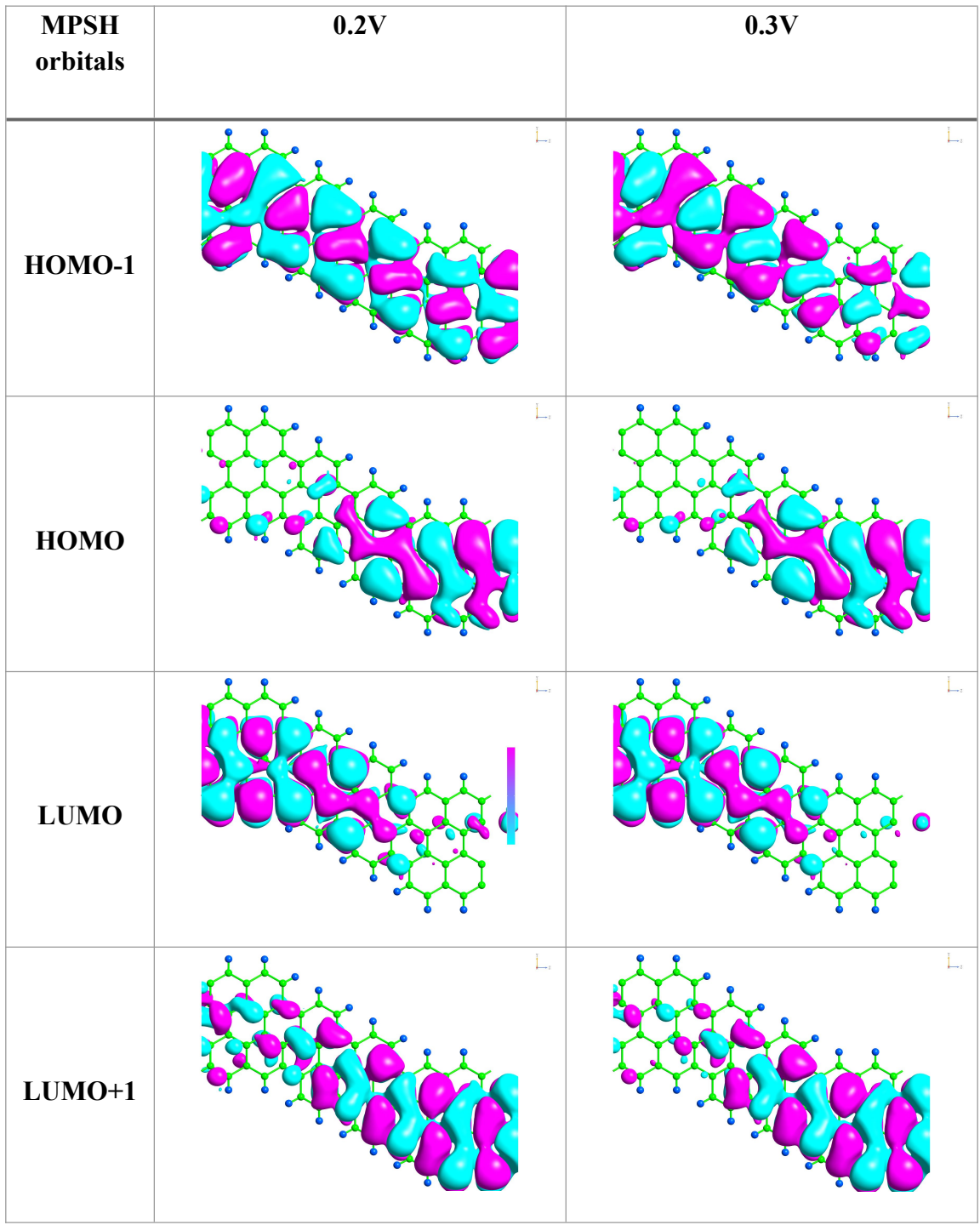

Fig. 6. Spatial distribution of MPSH HOMO-1, HOMO, and LUMO for 3-ZAZ SiNRs under different applied bias.

## 4. Conclusion

The transport properties of several ZAZ SiNRs junction are performed by using the density functional theory combined with nonequilibrium Green's function method.It is found that the differential conductances of the three systems decrease with an order of 5-ZAZ>4-ZAZ>3-ZAZ. Interestingly,the NDR phenomenon can be observed within certain bias voltage range only in 3-ZAZ SiNRs.It is suggested that the NDR behavior originates from the suppression of the frontier molecular orbitals LUMO of 3-ZAZ structure with the bias increasing.

## 5. Acknowledgments

The authors would like to acknowledge the support by the Project 61177076 supported by National Natural Science Foundation of China and the Fundamental Research Funds for the Central Universities of China, No. 2012-Ia-051.